\documentclass[12pt]{article}
\usepackage{jcappub}
\usepackage[top=1in, bottom=1in, left=1in, right=1in]{geometry}
 \pdfoutput=1
\usepackage[font=footnotesize]{caption}
\usepackage{array}
\usepackage{enumerate}
\usepackage{float}
\usepackage{subfig}
\usepackage{multicol}
\usepackage{multirow}
\usepackage{epsfig}
\usepackage{graphicx}
\usepackage{pict2e}
\usepackage{color}
\usepackage{amsmath,amsfonts,amssymb}
\usepackage{bbm}
\usepackage[utf8]{inputenc}
\usepackage[english]{babel}
\usepackage{xspace}

\newcommand{\be}{\begin{equation}}
\newcommand{\ee}{\end{equation}}
\newcommand{\bea}{\begin{eqnarray}}
\newcommand{\eea}{\end{eqnarray}}

\DeclareMathOperator{\tr}{\text{tr}}

\newcommand{\mpl}{M_{\text{P}}}

\title{Vacuum energy sequestering and conformal symmetry}

\begin{abstract}
{In a series of recent papers Kaloper and Padilla proposed a mechanism to sequester standard model vacuum contributions to the cosmological constant. We study the consequences of embedding their proposal into a fully local quantum theory. In the original work, the bare cosmological constant $\Lambda$ and a scaling parameter $\lambda$ are introduced as global fields. We find that in the local case the resulting Lagrangian is that of a spontaneously broken conformal field theory where $\lambda$ plays the role of the dilaton. A vanishing or a small cosmological constant is thus a consequence of the underlying conformal field theory structure.}
\end{abstract}

\author{Ido Ben-Dayan$^{a}$,
Robert Richter$^{b}$,
Fabian Ruehle$^{a}$,
Alexander Westphal$^{a}$}
\affiliation{${^a}$ Deutsches Elektronen-Synchrotron DESY, Theory Group, 22603 Hamburg, Germany\\
$^{b}$ II. Institut f\"ur Theoretische Physik, Hamburg University, Germany}
\emailAdd{ido.bendayan@desy.de, robert.richter@desy.de, fabian.ruehle@desy.de, alexander.westphal@desy.de}
\begin{document}

\begin{flushright}
\small
 DESY-15-119\\
 ZMP-HH/15-20
\end{flushright}
\bigskip
\bigskip\bigskip\bigskip\bigskip

\maketitle

\section{Introduction}
\label{sec:Introduction}

The cosmological constant (CC) problem is one of the most severe fine tuning problems of modern day physics \cite{Zeldovich:1967gd,Weinberg:1988cp}. Even when the vacuum energy contributions from quantum gravity are ignored, the quantum field theory (QFT) that describes the standard model of particle physics (SM) gives contributions to the vacuum energy that are huge compared to the observed CC.

In \cite{Kaloper:2013zca, Kaloper:2014dqa} the authors propose a mechanism that cancels the SM matter sector quantum corrections to the cosmological constant. For this mechanism to work two ingredients are crucial: first, two auxiliary fields $\lambda$, $\Lambda$ need to be added to the Lagrangian and second, the action needs to contain a function $\sigma$ which has a fixed dependence on these auxiliary fields and, more importantly, is outside of the spacetime integral. Finally,  one introduces an energy scale $\mu$ (which is non-physical) for dimensional reasons. The modified action then reads
\begin{align}
 \label{eq:KPAction}
 S=\int d^4x\sqrt{g}\left[\frac{\mpl^2}{2}R-\Lambda-\lambda^4\mathcal{L}(\lambda^{-2}g^{\mu\nu},\Phi)\right]+\sigma\left(\frac{\Lambda}{\lambda^4\mu^4}\right) \,.
\end{align}
Varying the action \eqref{eq:KPAction} w.r.t.\ $\Lambda$, $\lambda$ one obtains two equations which force the bare CC to be the "historic" average of the trace of the energy momentum tensor, $\Lambda=\langle T^{\mu}_{\mu}\rangle/4$, with $T_{\mu \nu}= \frac{-2}{\sqrt{g}} \frac{\delta S_m}{ \delta g^{\mu \nu}} $ where $\langle X \rangle$ is given by
\begin{align}
\langle X \rangle\equiv\frac{\int d^4x\sqrt{g}X}{\int d^4x \sqrt{g}}\,.
\end{align}
Using the latter, the Einstein equations become
\begin{align}
\label{eq:sequesteredEFE}
\frac{\mpl^2}{2}G^{\mu}_{\nu}=T^{\mu}_{\nu}-\frac{1}{4}\delta^{\mu}_{\nu}\langle T^{\alpha}_{\alpha}\rangle \,.
\end{align}
Thus in an old but finite universe the CC is very small and classical and quantum contributions arising from the SM are cancelled.

The decisive ingredient for this mechanism to work is that the auxiliary fields $\lambda$ and $\Lambda$ are global. This is enforced by introducing the function $\sigma$ containing the two auxiliary fields outside of the spacetime integral. As a consequence, the equation of motion for $\lambda$ and $\Lambda$ enforce a global rather than a local solution. This is crucial in order to ensure that the fields do not receive any contributions from quantum corrections: they are global constants and there is a sector (i.e.\ the $\sigma$-function) which only couples the two fields among themselves. Consequently they are neither subject to their own quantum corrections nor to quantum corrections of the other local, bona fide quantum fields in the theory.

While there is in principle nothing wrong with adding such a $\sigma$ term to the action, it is rather unconventional and does not arise from any known physics.\footnote{See, however \cite{Coleman:1988tj,Linde:1988ws,Tseytlin:1990hn,Davidson:2009mp,Davidson:2014vda}  for a discussion on the generation of such non-local terms in the context of globally interacting universes.}  Naively, one may think that such a term might arise from an instantonic sector of the theory for which the spacetime integral can be carried out explicitly. However, in deriving the effective theory one integrates out the instantonic configurations in the path integral and substitutes the result back into the action. In this way, one ends up again with a local theory.

In a more recent paper \cite{Kaloper:2015jra}, the authors discuss the implementation of the sequestering mechanism in a local theory, with the $\sigma$-function under the integral, but with a different measure. Their idea is to add an exact 4-form $F_4=d A_3$ coupled to the cosmological constant,
\begin{align}
 S\supset \int \Lambda \star\!1-\sigma\left(\frac\Lambda{\mu^4}\right)F_4\,.
\end{align}
The equation of motion for $A_3$ then dictates $\Lambda$ to be a constant.\footnote{Analogously, globality of $\lambda$ is enforced by introducing a second auxiliary 4-form field $G_4$, which was called $\tilde F_4$ in~\cite{Kaloper:2015jra}.} Again, this mechanism only works due to the global, auxiliary nature of $A_3$. More precisely, the action does not include a quadratic kinetic term $F_4\wedge\star F_4$ for $A_3$ but only the topological boundary term. However, making $A_3$ local by adding such a kinetic term modifies the equations of motion, rendering $\Lambda$ and $\lambda$ local quantum fields. As discussed in detail in the next section in such a scenario the sequestering mechanism fails due to quantum corrections.

Before we turn to the discussion of \emph{local} versus \emph{global} variables in the context of the sequestering mechanism, let us study the scaling dependence of the fields occurring in \eqref{eq:KPAction}.

\subsubsection*{Coupling to the Einstein--Hilbert term}

From  \eqref{eq:KPAction} one immediately determines that the inverse metric $g^{\mu\nu}$ scales as $\lambda^{-2}g^{\mu\nu}$, implying $g_{\mu\nu}$  scaling as $\lambda^{2}g_{\mu\nu}$. Consequently, the scaling of $\sqrt{g}=\sqrt{\det(g_{\mu\nu})}$ is given by $\lambda^4\sqrt{g_{\mu\nu}}$. The Riemann curvature tensor $R^{\rho}_{\mu\sigma\nu}=\partial_\nu (g^{\alpha\rho}(\partial_{\sigma}g_{\alpha\mu}+\ldots)+\ldots$ as well as the Ricci tensor $R_{\mu\nu}=R^{\rho}_{\mu\rho\nu}$ do not scale, while the scalar curvature $R=g^{\mu\nu}R_{\mu\nu}$ exhibits the scaling behavior $\lambda^{-2}R$.

\subsubsection*{Coupling to the bare CC}

Since $\Lambda$ is just a number, it does not involve any $\lambda$-dependence.

\subsubsection*{Coupling to the matter Lagrangian}
In order to derive the $\lambda$ scalings of the matter fields we investigate the canonically normalized kinetic terms. Starting with the kinetic terms for the gauge fields, $\tr (g^{\mu\rho}g^{\nu\sigma} F_{\mu\nu}F_{\rho\sigma})$ with $F=dA+A\wedge A$, we find that if $\mathcal{L}_m$ is to depend on $\lambda$ as $\lambda^{-4}$ to cancel the contribution from the integral measure, then $A$ does not scale with $\lambda$. 

From the bosonic kinetic terms $g^{\mu\nu}\partial_\mu\phi\partial_{\nu}\phi$ we find that the canonically normalized field is $\varphi=\lambda\phi$.

Next we look at the fermionic kinetic terms $\chi\gamma^\mu\partial_\mu\overline{\chi}$. From the Clifford algebra $\{\gamma^\mu,\gamma^\nu\}=2g^{\mu\nu}$ we find that $\gamma^\mu$ has to scale with  $\lambda$ as $\lambda^{-1}\gamma^{\mu}$ and hence the canonically normalized field is $\psi=\lambda^{3/2}\chi$.

With this combination, the trilinear Yukawa couplings $\phi\chi\overline\chi$ reads in terms of the canonically normalized fields $\lambda^{-4}\varphi\psi\overline\psi$, just as the other terms in $\mathcal{L}$. In this way, the field $\lambda$ sets the hierarchy between the matter scale and the Planck scale, 
\begin{align}
\label{eq:PhysicalMass}
\frac{m_\text{phys}}{\mpl}=\lambda\frac{m}{\mpl}\,,
\end{align}
where $m$ is the bare mass entering the Lagrangian $\mathcal{L}$ in \eqref{eq:KPAction} and $m_\text{phys}$ is the physical mass of the canonically normalized field $\varphi,\psi$, i.e.\
\begin{align}
\begin{split}
S_\text{m} 	&=\int d^4x\sqrt{g}\lambda^4[\lambda^{-2}g^{\mu\nu}\partial_\mu\phi\partial_{\nu}\overline{\phi}+\chi\lambda^{-1}\gamma^\mu\partial_\mu\overline{\chi}+\tr(\lambda^{-4}g^{\mu\rho}g^{\nu\sigma} F_{\mu\nu}F_{\rho\sigma})+m^2\phi^2+m\chi\overline{\chi}]\\
		&=\int d^4x\sqrt{g}[g^{\mu\nu}\partial_\mu\varphi\partial_{\nu}\overline{\varphi}+\psi\gamma^\mu\partial_\mu\overline{\psi}+\tr(g^{\mu\rho}g^{\nu\sigma} F_{\mu\nu}F_{\rho\sigma})+m_\text{phys}^2\varphi\overline{\varphi}+m_\text{phys}\psi\overline{\psi}]\,.
\end{split}
\end{align}

In summary we see that the different parts exhibit different scalings with $\lambda$: The scaling of the fields in the matter Lagrangian are such that they cancel the $\lambda^4$ scaling coming from the integral measure. In contrast, the Einstein--Hilbert term scales with $\lambda^{-2}$, and there is no $\lambda$-dependent prefactor in front of this term in~\eqref{eq:KPAction}. Consequently, the pure gravitational part of the action will not be sequestered. Higher loop quantum corrections will not spoil the sequestering by choosing a UV regulator and subtraction scale that have exactly the same $\lambda$ scaling, for instance by using Pauli-Villars regulators, \cite{Kaloper:2014dqa}. This means that the scaling symmetry
\begin{align}
 \lambda\rightarrow \Omega\lambda\,,\quad g_{\mu\nu}\rightarrow\Omega^{-2}g_{\mu\nu}\,,\qquad\Lambda\rightarrow\Omega^4\Lambda
\end{align}
is exact in the matter (and the bare cosmological constant) sector, but broken by the Einstein--Hilbert term. In this sense it is only an approximate symmetry and this serves (together with another scaling symmetry) to explain the smallness of the cosmological constant.

\section{Dropping the global \texorpdfstring{$\boldsymbol\sigma$}{sigma}-term -- dilaton effective action and spontaneously broken scale symmetry}

Let us start with the following action in four dimensions
\begin{align}
 \label{eq:NewAction}
 S=\int d^4 x \sqrt{g} \left[\frac{\mpl^2}{2}R - \lambda^{4}  \Lambda - \lambda^{4} \mathcal{L}_m(\lambda^{-2} g^{\mu\nu},\Phi)\right]\equiv\int d^4 x \sqrt{g} \left(\frac{\mpl^2}{2}R - \lambda^{4}  \Lambda\right)+S_m\,.
\end{align}
This is a slight modification of \eqref{eq:KPAction}: the global dynamical variable $\lambda$ is coupled to the CC $\Lambda$ and we did not include a $\sigma$-function outside the integral. $\mathcal{L}_m$ denotes the matter sector which scales with $\lambda$. We will see momentarily that the absence of the truly global $\sigma$-function will lead to a contradiction concerning the assumed globality of $\lambda$. Note that in contrast to the original mechanism proposed in  \cite{Kaloper:2013zca, Kaloper:2014dqa} the cosmological constant is not  promoted to a global dynamical variable.

Before studying the equations of motion (e.o.m.) let us first define the energy momentum tensor arising from the matter part and its corresponding trace: 
\begin{align}
\label{eq:Tmunu}
T_{\mu \nu}= \frac{-2}{\sqrt{g}} \frac{\delta S_m}{ \delta g^{\mu \nu}} = 2 \lambda^4 \left[ -\frac{1}{2} g_{\mu \nu} \,\,\mathcal{L} (\lambda^{-2} g^{\mu\nu} , \Phi) + \lambda^{-2} \frac{\delta \mathcal{L} (\lambda^{-2} g^{\mu\nu} , \Phi)}{ \delta (\lambda^{-2} g^{\mu \nu})} \right]\,,
\end{align}
where $S_m$ indicates the matter part of the action~\eqref{eq:NewAction}. The resulting trace is then given by
\begin{align}
\label{eq:Ttrace}
T^{\mu}_{ \mu}= 2 \lambda^4 \left[ -2 \, \mathcal{L} (\lambda^{-2} g^{\mu\nu} , \Phi) + \lambda^{-2}\, g^{\mu\nu}  \frac{\delta \mathcal{L} (\lambda^{-2} g^{\mu\nu} , \Phi)}{ \delta (\lambda^{-2} g^{\mu \nu})} \right]\,.
\end{align}

The e.o.m.\ arising from varying with respect to $\lambda$ gives
\begin{align}
\label{eq:VARlambda}
\frac{\delta S}{ \delta \lambda} =0 = \int d^4 x \sqrt{g} \left[ -4 \lambda^{3} \Lambda + 2 \lambda^{3} \left(-2\mathcal{L} (\lambda^{-2} g^{\mu\nu} , \Phi) + \lambda^{-2} \frac{\delta \mathcal{L} (\lambda^{-2} g^{\mu\nu} , \Phi)}{ \delta (\lambda^{-2} g^{\mu \nu})} \right)
  \right]\,.
\end{align}
\emph{Assuming} that $\lambda$ is a global field we have two possible e.o.m.\ arising from this variation. One may choose to pull $\lambda$ out of the spacetime integral and thus arrives at the \emph{global} version of the e.o.m.\ for $\lambda$,
\begin{align}
\label{eq:EOMglobal}
\lambda^4 \Lambda = \frac{1}{4} \frac{\int d^4 x \sqrt{g} T^{\mu}
_{\mu}}{ \int d^4 x \sqrt{g}} = \frac{1}{4} \langle  T^{\mu}
_{\mu} \rangle \,.
\end{align}
However, since all terms are inside the spacetime integral one can cancel the variation also locally, implying a \emph{local} e.o.m.\ for $\lambda$,
\begin{align}
\label{eq:EOMlocal}
\lambda^4 \Lambda =  \frac{1}{4}  T^{\mu}_{\mu}  \,.
\end{align}

Next, varying \eqref{eq:NewAction} with respect to the metric $g^{\mu\nu}$ one obtains
\begin{align}
\frac{\delta S}{\delta{g^{\mu\nu}}} = 0 = \frac{\mpl^2}{2 }\, \sqrt{g}\, G_{\mu\nu} + \frac{1}{2}\,\sqrt{g} \,g_{\mu \nu} \lambda^4 \Lambda  - \frac{1}{2}\,\sqrt{g}\, T_{\mu \nu}\,.
\end{align}
If one \emph{chooses} the \emph{global} e.o.m.\ \eqref{eq:EOMglobal} for $\lambda$ and substitutes it into the e.o.m.\ for the metric one recovers~\eqref{eq:sequesteredEFE},
\begin{align}
\label{eq:sequesteredEFE2}
\frac{\mpl^2}{2 }\,G_{\mu\nu} =T_\text{eff}\equiv T_{\mu \nu} -  \frac{1}{4}\, g_{\mu \nu} \langle T^{\alpha}_{ \alpha} \rangle\,.
\end{align}
Then averaging over this equation one obtains
\begin{align}\label{eq:avEOM}
\langle G^\mu_\mu\rangle=\langle \left(T_\text{eff}\right)^\mu_\mu\rangle\equiv \langle T^\mu_\mu -   \langle T^{\mu}_{ \mu} \rangle \rangle=0
\end{align}
for the pure matter sector contribution to the CC. However, we only ended up with the result~\eqref{eq:sequesteredEFE} and its sequestering of the matter sector CC due to the fact that we picked the global e.o.m.\ for $\lambda$ \emph{by hand}.\footnote{For an interesting discussion on incorporating global variables in a path integral, see \cite{Gabadadze:2014rwa}.} This result begs a question: there is no obstruction to $\lambda$ being a spacetime-dependent field that obeys all the rules of standard quantum field theory. As such, it will generate local equations of motion and be subject to renormalization. In the absence of an underlying symmetry there will be no control over such radiative corrections and the sequestering will be lost.\footnote{More explicitly, the sequestering cancellation in~\eqref{eq:avEOM} depends on the precise prefactor $1/4$ on the r.h.s.\ of~\eqref{eq:sequesteredEFE2} which in turn arises from the particular combination of the tree-level $\lambda$-scaling powers in the matter action. Since quantum corrections from all $\lambda$-coupled matter fields will change these scaling powers, the coefficient $1/4$ in~\eqref{eq:sequesteredEFE2} will get corrected, destroying the sequestering cancellation.} In the original mechanism \cite{Kaloper:2013zca, Kaloper:2014dqa}, the $\sigma$-function prohibited the existence of local solutions allowing only global ones, thus avoiding the standard quantization and renormalization procedure. As there is no truly global $\sigma$-function present in the action here, choosing a global $\lambda$ is ad-hoc and unjustified.\footnote{At this level, imposing e.o.m.\ for $\lambda$ is on the same footing as imposing  e.o.m.\ for the (Planck) mass, which is not done in conventional QFTs. To see this, absorb $\lambda$ into the metric, which gives the action in Jordan frame. The action is then $S=S_\text{EH}/\lambda^2+\hat S_m$, where $S_\text{EH}$ is the Einstein-Hilbert action and $\hat S_m=S_{\Lambda}+S_m$. Now $\lambda$ can be absorbed into the Planck mass to define the physical Planck mass $\mpl^\text{phys}=\mpl/\lambda$, such that a variation w.r.t.\ $\lambda$ is like a variation w.r.t.\ $\mpl^\text{phys}$ in another frame.}

For a solution to the matter sector CC, we hence desire an effective action for a dynamical field $\lambda$ coupled to a scaling matter sector with tree-level scaling powers as in the action~\eqref{eq:NewAction}, and a locally vanishing trace $T^\mu_\mu=0$. We can infer the structure of the action for $\lambda$ by observing that we can get the scaling powers of the matter sector by starting according to~\cite{Komargodski:2011vj} with an Einstein frame effective action in a spontaneously broken CFT containing the dilaton $\Sigma$ in the metric $\hat g_{\mu\nu}\equiv e^{2\ln\Sigma}g_{\mu\nu}$. Under \textit{global} Weyl transformations of the original metric
\begin{align}
\label{eq:WeylRescaledMetric}
 g_{\mu\nu}\to \lambda^2 g_{\mu\nu}\equiv e^{2\sigma}g_{\mu\nu}\,,
\end{align}
the metric $\hat g_{\mu\nu}$ is invariant provided that the dilaton shifts as
\begin{align}
\label{eq:DilatonShift}
 \ln\Sigma\to\ln\Sigma-\sigma\,.
\end{align}
The action that is invariant under global Weyl transformations then reads
\begin{align}
\label{eq:NewActionEinsteinFrame}
\tilde S=\int d^4 x \sqrt{\hat g} \left[\frac16 \hat R -\mathcal{L}_m(\hat g^{\mu\nu},\Phi)\right]\,.
\end{align}
For simplicity, we have absorbed a bare matter sector CC $\Lambda$ into the matter Lagrangian since they share the same tree-level $\lambda$-scaling power. 

If we allow for \textit{local} Weyl transformations, $\sigma=\sigma(x)$, we can use \eqref{eq:DilatonShift} to shift away the \textit{spacetime-dependent} field $\Sigma$, which removes the $\Sigma$-dependence from $\hat g_{\mu\nu}$. So let us review the behavior of \eqref{eq:NewActionEinsteinFrame} under local Weyl transformations \eqref{eq:WeylRescaledMetric}. The results can be found in textbooks, see e.g.\ Appendix D of \cite{Wald:1984rg}. We use $\nabla$ for the covariant derivative with the Christoffel connection $\Gamma$. Under the shift $g_{\mu\nu}\rightarrow \lambda^2 g_{\mu\nu}$, the Christoffel symbols shift as
\begin{align}
 \Gamma^\gamma_{\alpha\beta}\rightarrow \Gamma^\gamma_{\alpha\beta} \lambda^{-2}\frac12 g^{\gamma\delta}(g_{\beta\delta}\partial_\alpha\lambda^2+g_{\alpha\delta}\partial_\beta\lambda^2-g_{\alpha\beta}\partial_\delta\lambda^2)\,.
\end{align}
As a consequence, the Ricci tensor 
\begin{align}
R_{\mu\nu}=R^\rho_{\mu\rho\nu}=\partial_\rho\Gamma^\rho_{\nu\mu}-\partial_\nu\Gamma^\rho_{\rho\mu}+\Gamma^\rho_{\rho\alpha}\Gamma^\alpha_{\nu\mu}-\Gamma^\rho_{\nu\alpha}\Gamma^{\alpha}_{\rho\mu} 
\end{align}
shifts as
\begin{align}
\begin{split}
R_{\mu\nu} \rightarrow R_{\mu\nu} &- 2\nabla_\mu\nabla_\nu\ln\lambda-g_{\mu\nu}g^{\alpha\beta}\nabla_\alpha\nabla_\beta\ln \lambda+2(\nabla_\mu\ln\lambda)(\nabla_\nu\ln\lambda)\\
				  &-2g_{\mu\nu}g^{\alpha\beta}(\nabla_\alpha\ln\lambda)(\nabla_\beta\ln\lambda)\,, 
\end{split}
\end{align}
where we used that $g^{\mu\nu}g_{\mu\nu}=4$. Finally, the Ricci scalar $R=g^{\mu\nu}R_{\mu\nu}$ shifts as
\begin{align}
\label{eq:RicciScalarShift}
 R\rightarrow\lambda^{-2}[R-6g^{\alpha\beta}\nabla_\alpha\nabla_\beta\ln\lambda-6g^{\alpha\beta}(\nabla_\alpha\ln\lambda)(\nabla_\beta\ln\lambda)]\,.
\end{align}
We now insert this in the action \eqref{eq:NewActionEinsteinFrame} whose integral measure transforms as $\sqrt{g}\rightarrow\lambda^{4}\sqrt{g}$. As a last step we perform an integration by parts to get rid of the $\nabla_\alpha\nabla^\alpha\ln\lambda$ term in \eqref{eq:RicciScalarShift}. Using the identities
\begin{align}
 \nabla_\alpha\ln\lambda=\partial_\alpha\ln\lambda\,,\quad\nabla_\alpha\nabla^\alpha\ln\lambda=(\partial_\alpha+\Gamma^\gamma_{\gamma\alpha})\partial^\alpha\ln\lambda\,,\quad\partial_\alpha\sqrt{g}=\sqrt{g}\,\Gamma^\gamma_{\gamma\alpha}\,,
\end{align}
the Christoffel connection cancels against the derivative of the measure, such that we arrive at the Jordan frame action

\begin{align}
\label{eq:NewAction2}
\begin{split}
S&=\int d^4 x \sqrt{g} \left[\frac{\lambda^2}{6} R+ \lambda^2(\partial \ln\lambda)^2-\lambda^4\mathcal{L}_m(\lambda^{-2}g^{\mu\nu},\Phi)\right]\\
&=\int d^4 x \sqrt{g} \left[\frac{\lambda^2}{6} R+ (\partial \lambda)^2-\lambda^4\mathcal{L}_m(\lambda^{-2}g^{\mu\nu},\Phi)\right]\,.
\end{split}
\end{align}
The appearance of $\lambda$ takes the precise form of the effective action of a theory with scale invariance broken at some scale $f$ given by $\langle\lambda\rangle$.\footnote{We have absorbed the UV mass scale $\hat M$ implicitly present in the action eq.~\eqref{eq:NewActionEinsteinFrame} into our definition of $\lambda$ here.} Note that the crucial difference to the theory given by eq.~\eqref{eq:NewAction} arises by demanding $\lambda=\lambda(x)$ to be a local and dynamical quantity implying that local Weyl transformations have to act on the metric that $R$ depends on as well. Otherwise $\lambda$ would not acquire a kinetic term and could not become a part of a fully local QFT, i.e.\ of a spontaneously broken CFT. In such a theory there is the Goldstone boson of spontaneous scale invariance breaking, namely the dilaton $\Sigma$, and scale invariance determines its effective action below the scale $f$. Indeed, we can choose a gauge where $\sigma=\ln\Sigma$, thus absorbing the dilaton via \eqref{eq:DilatonShift} into the Weyl transformation of the metric. In this gauge we can write $\lambda=\langle\lambda\rangle\Sigma=f\Sigma$ according to \eqref{eq:WeylRescaledMetric}. The resulting effective action below the scale $f$ then reads
\begin{align}
\label{eq:DilatonAction}
S&=f^2\int d^4 x \sqrt{g} \left[\frac{\Sigma^2}{6} R+ (\partial \Sigma)^2-\kappa f^2\Sigma^4\right]\,.
\end{align}
This action is invariant under scale transformations of the metric $g_{\mu\nu}\to \Omega^2g_{\mu\nu}$ if at the same time mass scales like $f$ transform as $f\to \Omega^{-1} f$ and the associated canonical dilaton transforms as $\phi=f\Sigma \to\Omega^{-1}\phi$. By comparing with~\eqref{eq:NewAction2}, we see that 
$\lambda$ is the dilaton of spontaneously broken scale invariance.

This match is not accidental for the following reason. We started above by demanding a local action with local invariance under $\lambda$-scalings and demanding $T^\mu_\mu=0$. One can show that any unitary QFT with conserved $T_{\mu\nu}$ which also has vanishing trace $T^\mu_\mu=0$ possesses a local scale symmetry with the conserved scale transformation current $j_{\mu}\sim T_{\mu\nu}X^\nu$ \cite{Dymarsky:2013pqa,Nakayama:2013is}. Hence, under our assumptions the theory must have at most spontaneously broken scale invariance which forces it to take the form discussed above.

Moreover, it has been proven in recent years that any 4D Poincar\'e invariant unitary QFT with scale invariance has full conformal symmetry at the perturbative level \cite{Komargodski:2011vj, Luty:2012ww, Dymarsky:2013pqa,Nakayama:2013is}. Therefore, under our assumptions removing the manifestly global $\sigma$-function forces the complete matter sector of the theory to take the form of a conformal field theory (CFT) with spontaneously broken conformal and scale invariance.

Finally, we note that in a scale invariant or conformal field theory the scaling of $f\to \Omega^{-1} f$ or correspondingly $\lambda\to\Omega^{-1}\lambda$ under a scale transformations of the metric forbids any potential terms except for the quartic self-coupling and in particular a cosmological constant term. Hence, all quantum corrections to the vacuum energy of the theory can be absorbed into renormalizing the quartic self-coupling. In the absence of gravity, the dilaton always runs to zero for $\kappa>0$, restoring conformality. For $\kappa<0$ the dilaton runs away to infinity and the theory is ill-defined. So the spontaneously broken phase requires a tuned $\kappa=0$.\footnote{Attempts to avoid or explain such a tuning exist in a non-normalizable case \cite{Shaposhnikov:2008xi} and from a five-dimensional perspective in \cite{Coradeschi:2013gda, Bellazzini:2013fga}. Furthermore, see the elaborate analysis of \cite{tHooft:2011aa}.} This implies the well-known statement that in a spontaneously broken CFT or scale invariant theory the full quantum-corrected CFT sector cosmological constant vanishes \emph{in the absence of gravity} \cite{Berman:2002kd}. In the presence of gravity (which we take to be non-conformal; at any rate no quantum-conformal field theory of gravity is known above two dimensions) this situation changes due to the conformal anomaly \cite{Cardy:1988cwa,Antoniadis:1991fa,Komargodski:2011vj}. Coupling the CFT to gravity causes the anomaly to dictate a finite contribution to $T^\mu_\mu$ from all CFT vacuum energy diagrams \emph{without} internal graviton lines which scales like
\begin{align}
\label{eq:anomaly}
T^\mu_\mu={\cal O}(R^2)\,,
\end{align}
where $R^2$ symbolizes the various quadratic invariants built from curvature tensors. Hence, if all contributions to the vacuum energy are controlled by the anomaly, the vacuum energy is parametrically small for large universes.

Note that once scale invariance is broken, we expect the dilaton to acquire a mass from quantum effects of the gravitational sector, since all quantum field theories with GR as their low-energy limit constructed so far seem to break conformal symmetry at the quantum level (there are no quantum conformal theories of gravity known). Moreover, the role of Goldstone's theorem in ensuring the massless of the dilaton is somewhat different with respect to scale invariance, that is, a non-compact local symmetry, rather than a  \emph{compact global} internal symmetry. There are mechanisms that can ensure a naturally light dilaton, see e.g.\ \cite{Coradeschi:2013gda, Bellazzini:2013fga}. So for the dilaton to remain light, one has to apply such a mechanism or some other, yet unknown, mechanism, that keeps the dilaton naturally light.

The cosmon scenario \cite{Wetterich:1987fk,Wetterich:1987fm,Wetterich:2008sx} is also based on 'dilatation symmetry' and the generation of scales via the conformal anomaly. Let us reiterate the differences between the cosmon scenario and our approach. In the cosmon scenario, both the Planck mass and the cosmological constant are vevs of some scalar field in Brans-Dicke theory with a 'dilaton/cosmon' and a Higgs doublet as a starting point \cite{Wetterich:1987fm}. The outcome is that particle masses are time dependent, and the anomaly responsible for the generation of scales depends on the cosmon and the Higgs field. This is certainly a much more advanced program than our proposal. It also addresses many fundamental questions about the CC and even allows for a viable quintessence model \cite{Wetterich:2008sx}.  
 In our case, by limiting ourselves to the matter sector only, our Planck mass is not an induced quantity, and therefore, we cannot address the full CC problem.   We therefore do not have a prediction for the dilaton mass. On the other hand, the masses in our scenario will not be time-dependent, and the anomaly is given by gravitational anomaly only, i.e. schematically by terms like eq.~\eqref{eq:anomaly}.

Let us  make a final remark on the approach discussed in \cite{Kaloper:2015jra}. There the authors use  non-dynamical auxiliary 4-form field strengths $F_4$ and $G_4$ to enforce the global nature of $\Lambda$ and $\lambda$. However, making the auxiliary 4-form field strengths dynamical by inclusion of their kinetic terms $F_4\wedge \star F_4$ and $G_4\wedge \star G_4$, respectively, this model maps into the same class of effective actions of spontaneously broken scale-invariant theories.  To see this, it serves to note that in four dimensions upon including the kinetic term the solution for $G_4$ from Poincar{\'e} invariance is $G_4 = \phi(x)\epsilon_{\mu\nu\rho\sigma}dx^\mu dx^\nu dx^\rho dx^\sigma$, which introduces a dynamical field $\phi$ that becomes effectively the dilaton of spontaneous scale symmetry breaking.

\section{Conclusions}
\label{sec:conclusions}
We studied a local version of the sequestering mechanism for the vacuum energy as proposed in \cite{Kaloper:2013zca,Kaloper:2014dqa,Kaloper:2015jra}. As a first approach we treated the field $\lambda$, whose e.o.m.\ lead to the sequestering mechanism, as a global field but disposed of a global function $\sigma$ which was included in the original mechanism. We argue that this approach is inconsistent within the standard rules of quantum field theories.

We then investigate the case where $\lambda$ is fully local, i.e.\ where we take the local e.o.m.\ of $\lambda$ and add a kinetic term. In this case $\lambda$ plays the role of a dilaton and the action takes the form of a 4D CFT. In such theories the cosmological constant stays zero even after spontaneously breaking the CFT (or more precisely the scaling symmetry). Hence, any reasonable, local version of the global sequestering mechanism put forward in~\cite{Kaloper:2013zca, Kaloper:2014dqa} will lead to the matter sector of such a theory taking the form of spontaneously broken CFT. The matter sector vacuum energy is then no longer fully sequestered, but can be small by virtue of being controlled by the conformal anomaly and thus occurs at higher order in the background curvature. 

Phenomenologically, all such theories predict a massless or very light dilaton scalar degree of freedom which we do not observe and imply a very large breaking scale $f$. It is an open question whether further symmetry breaking patterns can be invoked to give the dilaton a mass \emph{without} destroying the control of the matter sector vacuum energy by the conformal anomaly. Furthermore, it remains to be seen whether the scales of the SM can be made consistent with the scale of the CFT breaking.

\section*{Acknowledgments}
We thank Rutger Boels, Wilfried Buchm\"uller, Aharon Davidson, Markus Dierigl, Gregory Gabadadze, Nemanja Kaloper, Jared Kaplan, Zohar Komargodski, Thomas Konstandin, Pedro Liendo, Markus Luty, Eliezer Rabinovici, and Marco Serone for useful discussions. This work was supported by the German Science Foundation (DFG) within the Collaborative Research Center (SFB) 676 ``Particles, Strings and the Early Universe''. AW and IBD are supported by the Impuls und Vernetzungsfond of the Helmholtz Association of German Research Centres under grant HZ-NG-603. AW acknowledges the support in part by National Science Foundation Grant No.~PHYS-1066293 and the stimulating and warm hospitality of the Aspen Center for Physics during the final phase of this work. Finally, RR acknowledges the support from the FP7 Marie Curie Actions of the European Commission, via the Intra-European Fellowships (project number: 328170).

\begin{footnotesize}
\providecommand{\href}[2]{#2}\end{footnotesize}
\end{document}